\documentclass{elsart}
\usepackage{graphicx}
\usepackage{amssymb}

\journal{New Astronomy Reviews}

\begin{document}

\begin{frontmatter}



\title{Applications of Gas Imaging Micro-Well Detectors to an Advanced Compton Telescope}

\thanks[nrc]{Supported by a National Research Council Research Associateship at GSFC}

\thanks[miller]{Current address: Department of Physics, University of Alabama in Huntsville, Huntsville, AL 35899, USA}


\author[GSFC]{P. F. Bloser\thanksref{nrc}},
\ead{bloser@milkyway.gsfc.nasa.gov}
\author[GSFC]{S. D. Hunter},
\author[UNH]{J. M. Ryan},
\author[UNH]{M. L. McConnell},
\author[UNH]{R. S. Miller\thanksref{miller}},
\author[PSU]{T. N. Jackson},
\author[PSU]{B. Bai},
\author[PSU]{S. Jung}

\address[GSFC]{NASA Goddard Space Flight Center, Code 661, Greenbelt, MD 20771, USA}
\address[UNH]{Space Science Center, University of New Hampshire, Durham, NH, 03824, USA}
\address[PSU]{Center for Thin Film Devices, Pennsylvania State University, University Park, PA 16802, USA}

\begin{abstract}
% Text of abstract
We present a concept for an Advanced Compton Telescope (ACT) based on the
use of pixelized gas micro-well detectors to form a three-dimensional
electron track imager.  A micro-well detector consists of an array of
individual micro-patterned proportional counters opposite a planar drift
electrode.  When combined with thin film transistor array readouts,
large gas volumes may be imaged with very good spatial and energy
resolution at reasonable cost.  The third dimension is determined from
the drift time of the ionization electrons. The primary advantage of
this approach is the excellent tracking of the Compton recoil electron
that is possible in a gas volume.  Such good electron tracking allows us to
reduce the point spread function of a single incident photon dramatically,
greatly improving the imaging capability and sensitivity.  The
polarization sensitivity, which relies on events with large Compton
scattering angles, is particularly enhanced.  We describe a possible ACT
implementation of this technique, in which the gas tracking volume is
surrounded by a CsI calorimeter, and present our plans to build and test a
small prototype over the next three years.

\end{abstract}

\begin{keyword}
gamma rays: observations \sep instrumentation: detectors 
\sep space vehicles: instruments \sep techniques: high angular resolution
\PACS  95.55.Ka \sep 29.40.Cs \sep 29.40.Gx
\end{keyword}
\end{frontmatter}

% main text

\section{Introduction}
\label{sect:intro}

The Advanced Compton Telescope (ACT) is envisioned as a $\sim 100$-fold increase
in sensitivity in the medium energy gamma-ray band (0.4-50 MeV) over CGRO/COMPTEL, the 
only Compton telescope flown in space with
enough sensitivity to make astronomical observations (Sch\"onfelder et al. 1993).
Many technological approaches are currently being studied to make ACT a reality
(Kurfess \& Kroeger 2001, and references therein).  Among these, one scheme, 
being pursued by both the MEGA (Kanbach et al. 2003) and TIGRE (O'Neill et al. 2003)
projects, seeks to increase sensitivity by tracking the recoil electron from the 
initial Compton scatter interaction.  The telescope consists of a tracker, made of
a stack of thin silicon strip detectors, and a calorimeter, made of small CsI
crystals.  The Compton interaction takes place in the tracker; the recoil electron
may be tracked through several layers of silicon, while the scattered photon is
absorbed in the calorimeter.  Measuring the electron momentum has the effect of 
reducing the typical 
Compton ``event circle'' to a smaller ``event arc,'' reducing background and
improving imaging.  We present a concept for ACT that 
takes electron tracking
one step further: a gas electron-track imager, instrumented with pixelized
micro-well detectors.

\section{Pixelized Micro-Well Detectors}
\label{sect:pmwd}

\begin{figure}[t]
\begin{minipage}[t]{6.75cm}
\includegraphics[height=6.4cm,width=6cm]{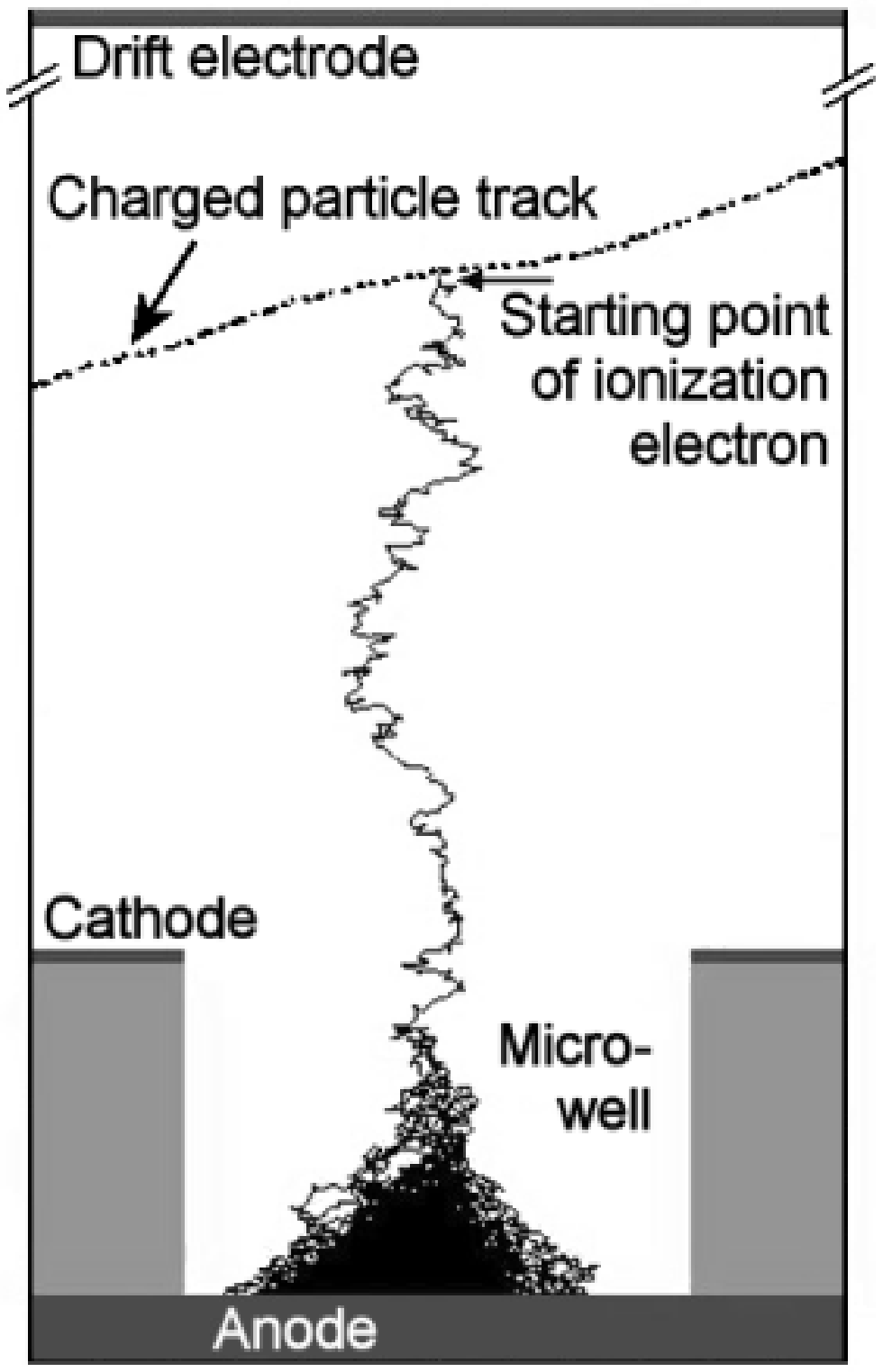}
\caption{Schematic (not to scale) of a micro-well detector.  Charge amplifcation
takes place in a high electric field inside the well.}
\label{fig:well}
\end{minipage}
\hspace*{0.3cm}
\begin{minipage}[t]{6.75cm}
\begin{center}
\includegraphics[height=6.4cm,width=6.74cm]{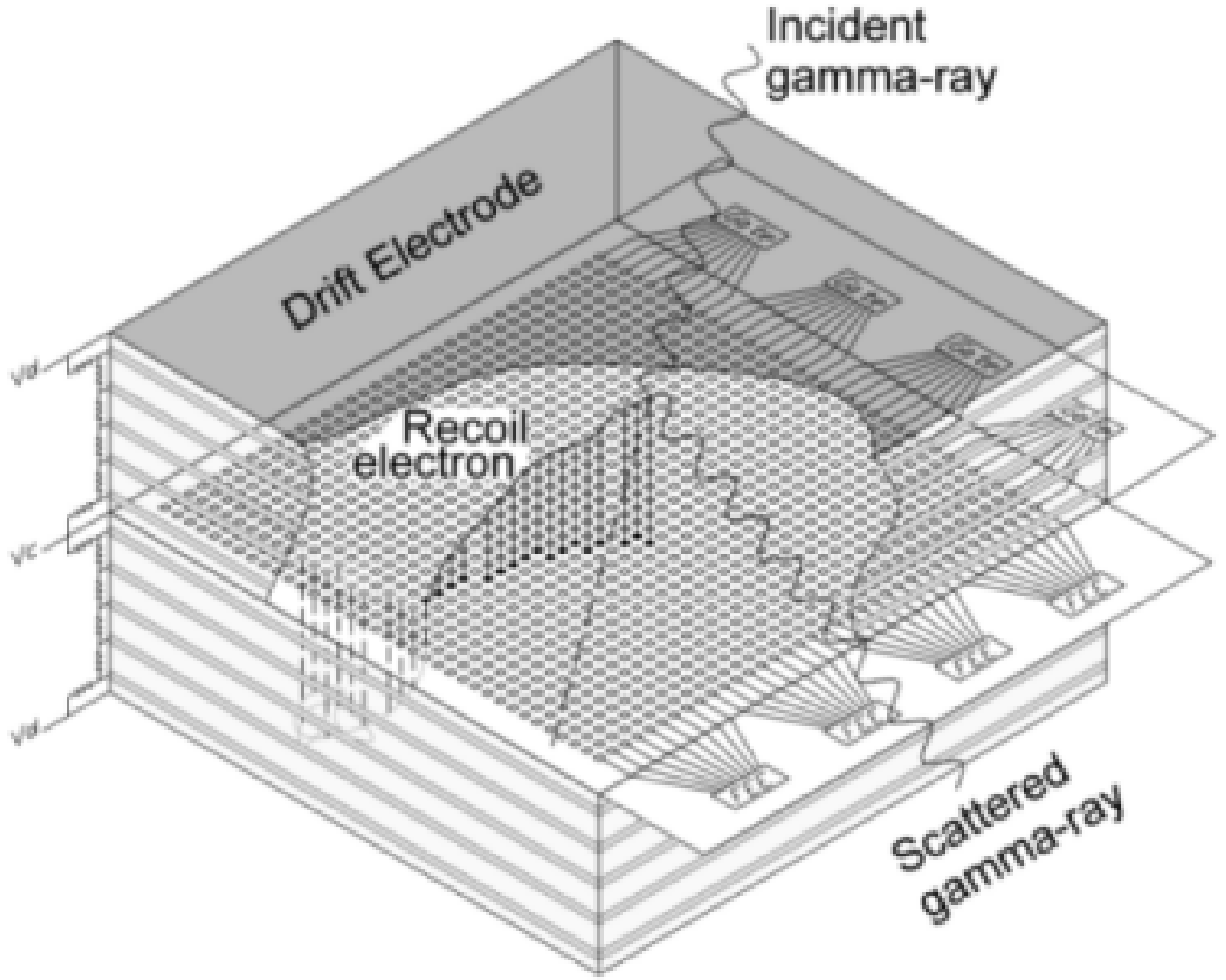}
\end{center}
\caption{TTID module consisting of two back-to-back PMWD-TFT arrays.  The track of
the recoil electron is imaged by the arrays, and the third dimension is calculated
from the charge drift time.}
\label{fig:ttid}
\end{minipage}
\end{figure}

The micro-well detector is a type of gas proportional counter based on micro-patterned
electrodes (Deines-Jones et al. 2002a, 2002b).  Each sensing element consists of a
charge-amplifying well (Figure~\ref{fig:well}).
The cathode and anode electrodes are deposited on opposite sides of an insulating substrate.
The well is formed as a cylindrical hole through the cathode and substrate, exposing the anode.
An array of such wells forms a detector, with the active tracking volume bounded by a drift
electrode.  Ionization electrons produced by the passage of
a fast charged particle drift toward the anodes and into the wells.  An ionization avalanche occurs
in each well, where an intense electric field is set up by the voltage applied between the anode 
and cathode.  Micro-well technology is very robust, and allows large
areas to be read out with good spatial ($\sim 100$ $\mu$m) and energy (18\% FWHM at 6 keV) 
resolution at low cost.

We are working to develop pixelized micro-well detectors (PMWDs) to enable true imaging of
charged particle tracks.  In this approach, each anode pad is connected to an element of a
thin-film transistor (TFT) array.  The individual transistor gates are connected in columns, 
and the outputs are connected in
rows.  The gate drivers for each column are then activated sequentially, allowing the charge collected
on the anode pads to be read out by charge-intergating amplifiers at the end of each row.  Thus a 
two-dimensional projected image of the charged particle track is recorded.  The third dimension may
be determined by measuring the drift time of the ionization electrons.  These combined PMWD-TFT arrays
will be assembled into modular detector units called three-dimensional track imaging detectors
(TTIDs), as shown in Figure~\ref{fig:ttid}.  Each TTID comprises two back-to-back PMWD-TFT arrays bounded
by drift electrodes and field-shaping electrodes on the four walls.  The electronics are distributed 
around the periphery of the module and then folded up along the walls.  

Our concept for a Compton telescope consists of a tracker made up of TTID modules surrounded by a 
calorimeter similar to that envisioned for MEGA or TIGRE.  The PMWDs would have a pitch of
$\sim 150$--200 $\mu$m, and the tracker would be filled with xenon gas at $\sim 3$ atm pressure.  These
numbers represent a compromise between high interaction probability, low electron scattering, 
high spatial resolution, and the number of electronics channels required.  The UV scintillation
light from the xenon serves as a trigger, both for the drift timing electronics and for 
coincidence with the calorimeter.

\section{Advantages of Electron Tracking for ACT}
\label{sect:tracking}

The primary advantage of a gas tracker for a Compton telescope is that the recoil electron
may be tracked far more accurately than in any other detector medium.  This is due to the 
low Coulomb scattering per track measurement distance in a uniform
low-density environment.  For xenon gas at 3 atm with a readout pitch of 150 $\mu$m, 
$\sim 100$ samples of the track may be recorded in $3.2 \times 10^{-3}$ radiation lengths.
This is the same number of radiation lengths as 300 $\mu$m of silicon, or one layer
of a typical silicon tracker as in MEGA or TIGRE, in which only one sample is possible.
Monte Carlo simulations made with Geant4 indicate that in a TTID with the above parameters
the initial direction of a 1 MeV electron may be determined to $\sim 7^{\circ}$ (rms), 
compared to $\sim 25^{\circ}$ expected for a silicon tracker.  (The effects of passive materials
are not yet included in these simulations or estimates.)

\begin{figure}
\begin{center}
\begin{tabular}{lr}
\includegraphics[height=6cm,width=6.7cm]{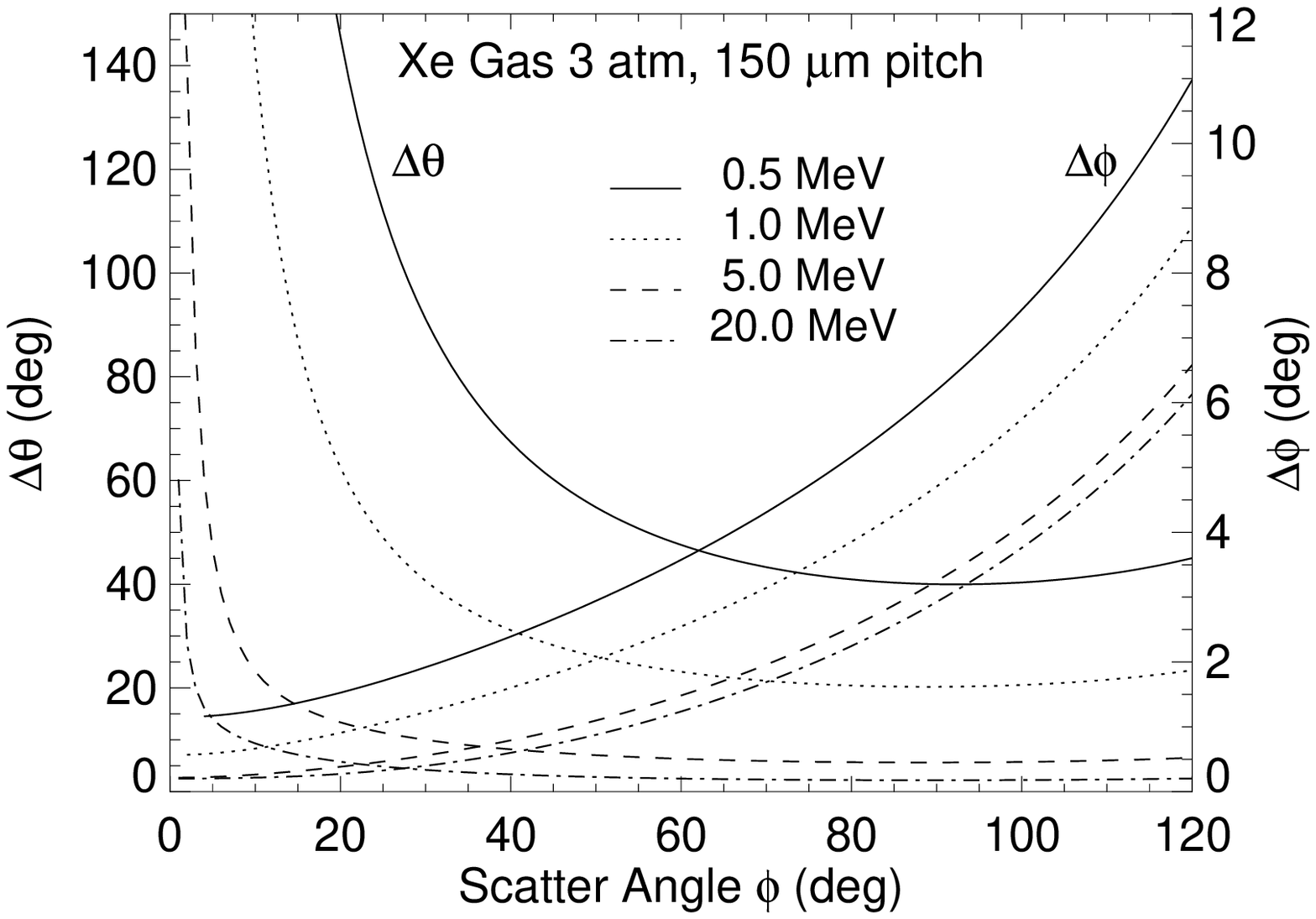}
& \includegraphics[height=6cm,width=6.7cm]{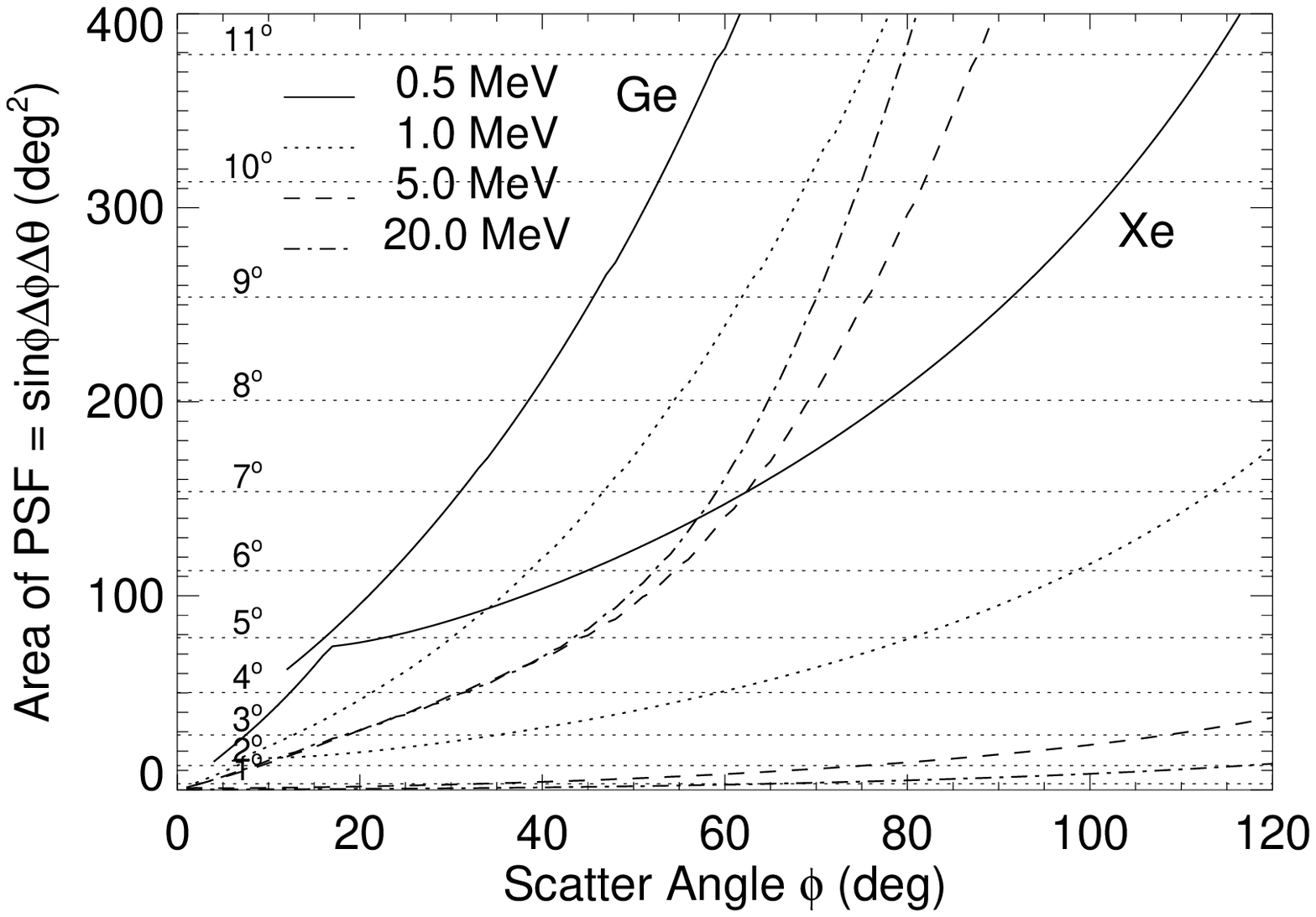}
\end{tabular}
\end{center}
\caption[example]
{ \label{fig:psf}
{\em Left:} Width $\Delta\phi$ and length $\Delta\theta$ (FWHM) of event arc for a xenon Compton
telescope as a function of Compton scatter angle $\phi$.
{\em Right:} Total angular area of the PSF for a xenon Compton telescope, compared with that of
a Ge triple-Compton telescope.  The horizontal lines indicate the equivalent angular radius.
}
\end{figure}

The primary advantage of electron tracking is the reduction of the point spread function (PSF).
There are two components to the PSF of a Compton telescope.  The first is 
the width of the event arc, equivalent to the error in the computed scatter angle
$\Delta\phi$.  (This is often referred to as the angular resolution meaure, or ARM.)  This
width is determined by the spatial and energy resolution of the tracker and calorimeter.  The
second component is the length of the event arc, $\Delta\theta$.  This is roughly given by the
error in the recoil electron's initial direction, projected onto the plane perpendicular to
the scattered photon direction.  We plot analytical calculations of these quantities for 
a xenon Compton telescope with
the parameters given above in Figure~\ref{fig:psf} (left) as a function of incident photon 
energy and Compton scatter angle $\phi$.
A CsI calorimeter with 5 mm $\times$ 5 mm pixels and an energy resolution of 5\% (FWHM) at 662 keV 
is assumed, as for MEGA and TIGRE.
The calculated quantities are for ``good'' events, in which the incident photon scatters once in the
xenon tracker and is then totally absorbed in the CsI.  We assume that the micro-well energy resolution
scales as the inverse square root of the energy (giving, for example, 1.5\% FWHM at 1 MeV) and that
the average distance between tracker and calorimeter interactions is 1 m.  The accuracy of
the location of the initial Compton scatter is on the order of the pitch and is neglected compared to
the 5 mm pixel size of the CsI.
The errors in the determination of the recoil electron direction are taken from Geant4 simulations.  
For low scatter angles
the electron receives very little energy, making tracking difficult, and so $\Delta\theta$ is large.
The calculation of $\Delta\phi$ includes an analytic correction for Doppler broadening 
(Zoglauer \& Kanbach 2003).  

The total angular area of the PSF is given by $A = \sin\phi\Delta\phi\Delta\theta$.  This is the region of
the sky from which the incident photon could have originated.  To decrease background
and improve imaging, it is critical to minimize this area.  Figure~\ref{fig:psf} (right) shows the
total PSF area for a Xe Compton telescope.  For comparison, the equivalent curves for a different ACT
scheme, the germanium triple-Compton approach (Kurfess \& Kroeger 2001), are shown as well.  The 
triple-Compton technique relies on detectors with excellent energy resolution to reconstruct multiple
(at least three) Compton scatter sequences so as to derive a photon's energy without the need for a
calorimeter.  For our calculations we assume that the second scatter occurs at the most likely
angle from the Klein-Nishina formula.  The thick Ge detectors assumed in Figure~\ref{fig:psf} achieve 
a very small $\Delta\phi$
due to their energy resolution, but are not able to track the recoil electron ($\Delta\theta = 2\pi$).
Therefore the PSF area is much larger, especially at large scatter angles.  The inclusion of large
scatter angles is necessary to achieve a large effective area.

\begin{figure}[t]
\begin{minipage}[t]{6.75cm}
\includegraphics[height=6cm,width=6.7cm]{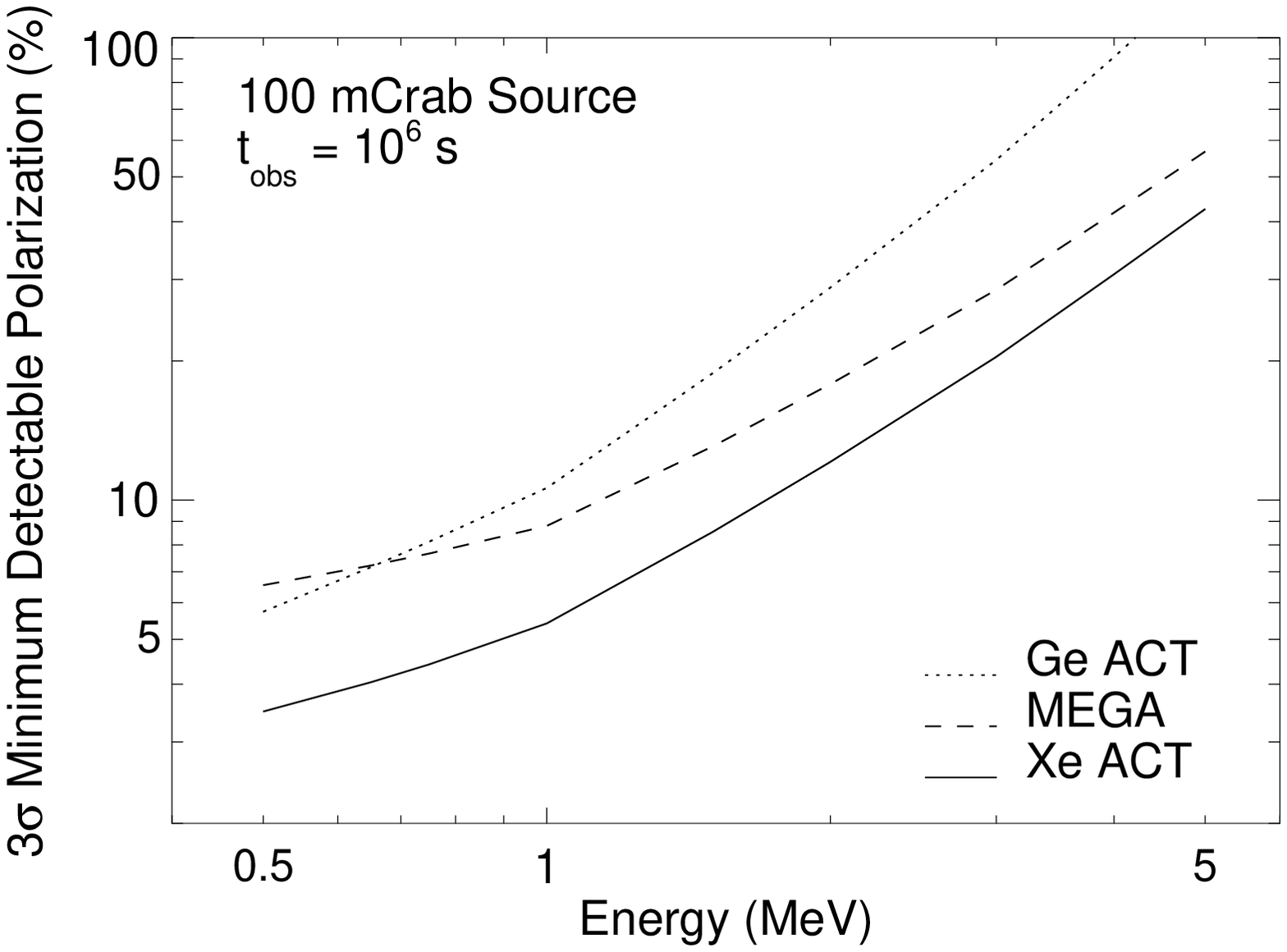}
\caption{Comparison of the minimum detectable polarization for an ACT based on the Ge triple-Compton,
MEGA, and Xe designs.  The same effective area, background, and azimuthal modulation are assumed for each.}
\label{fig:pol}
\end{minipage}
\hspace*{0.3cm}
\begin{minipage}[t]{6.75cm}
\begin{center}
\includegraphics[height=6cm,width=6.7cm]{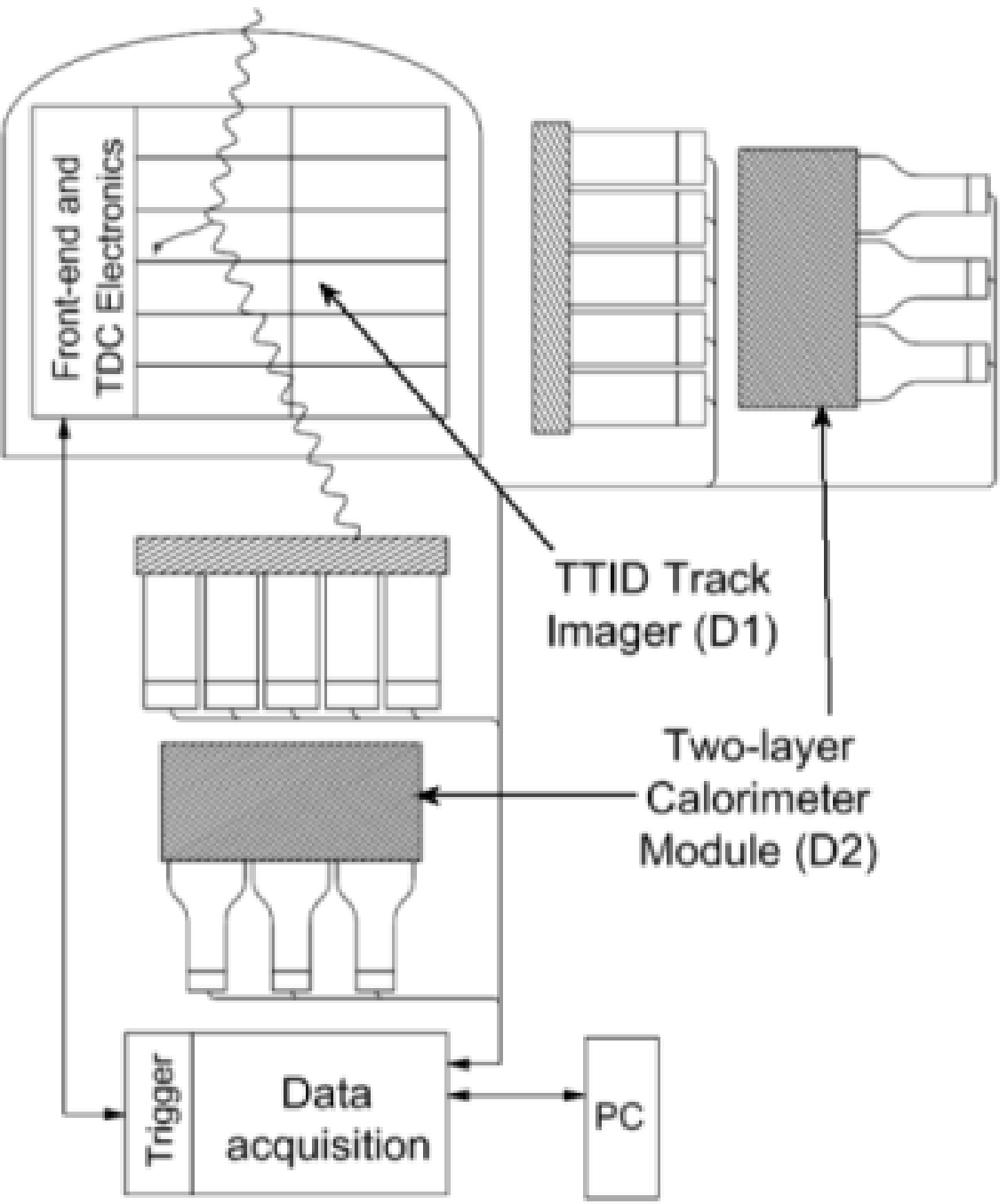}
\end{center}
\caption{
Block diagram of planned Xe Compton telescope prototype, showing Xe tracker, NaI caloimeter modules, 
and data aquisition.  The prototype is intended for a gamma-ray beam test in 2006.
}
\label{fig:prototype}
\end{minipage}
\end{figure}

Polarization sensitivity is a major goal of ACT.  Polarization is detected by measuring the
asymmetric azimuthal distribution of Compton-scattered polarized photons (Lei et al. 1997).  
Since this azimuthal asymmetry is most pronounced for large scatter angles, a Compton polarimeter
will benefit greatly from a low background for large-scatter angle events.  Figure~\ref{fig:pol}
compares the minimum detectable polarization (Lei et al. 1997) for ACT telescopes similar to 
the Ge triple-Compton, MEGA, and Xe designs for a $10^6$ s observation of a 100 mCrab source.
An effective area of 3000 cm$^2$ (typical of current ACT concepts; Kurfess \& Kroeger 2001) 
and a total background twice that of COMPTEL are assumed for each,
as well as the maximum possible azimuthal asymmetry,
so that only the effects of the PSF area are compared.  The Xe ACT concept achieves 2-3 times the
polarization sensitivity of the other designs due to its tracking ability.

Good electron tracking is also of great value for background rejection.  Recording the recoil electron
direction of motion allows the suppression of upward-moving background gamma rays from the atmosphere
(e.g., O'Neill et al. 2003).  In a gas detector, this technique will work well below 1 MeV.  In addition,
the complete kinematic information gained from the electron's motion permits consistency checks on the
reconstructed photon's direction and energy.  For example, events that are not fully absorbed in the 
calorimeter may be identified.  We calculate that incompletely absorbed 20 MeV photons may be
rejected for scattering angles greater than $10^{\circ}$.  

\section{Xe ACT Concept and Prototype Development}
\label{sect:act}

The low density of the gas tracker (the $1/e$ attenuation length for Compton scattering of 
1 MeV photons in 
3 atm of xenon gas is 11.6 m) implies that a large physical volume will be required to
achieve a high effective area.  
Initial Geant4 simulations indicate that a sensitivity $\sim 100$ times that of COMPTEL may be
achieved for a xenon tracking volume 2.5 m deep (0.22 attenuation lengths, similar to the MEGA
concept) $\times$ 2.5 m in diameter, surrounded by
CsI 10 cm thick on the bottom and 5 cm thick on the sides.  
This concept requires $\sim 200$ kg of Xe and $\sim 5800$ kg of CsI, a ratio of tracker to calorimeter
masses that is similar to that in the MEGA or TIGRE concepts.
The effective area for single Compton
scattering in the tracker followed by complete absorption in the calorimeter is $\sim 6700$ cm$^2$ 
at 1 MeV.  No passive material was included in
the simulation.  
Such a telescope would imply a total 
payload mass of $\sim 8000$ kg, well within the low earth orbit capability of launchers such as 
the Delta IV.  Thus the Xe telescope concept is feasible and meets the goals of ACT.

To demonstrate the advantages of the Xe ACT, we plan to build and test a prototype over the next
three years (Figure~\ref{fig:prototype}).  The PWMD-TFT arrays will be produced by Penn State.  
The tracker and electronics will
be designed and constructed at GSFC, while UNH will supply three simple NaI calorimeter modules
to be placed behind and on two sides of the tracker.  
Special attention will be paid to implementing the high density of channels and interconnects
and to minimizing the amount of passive material in the gas volume.
Our goal is to
test this prototype with 100\% polarized gamma rays from 0.7--100 MeV at Duke University's 
High Intensity Gamma Source in 2006 and to compare the results with more detailed simulations to prove
the value of precise electron tracking for ACT.

\end{document}